\documentclass[11pt]{article}

\usepackage{hyperref}
\usepackage{a4wide}
\usepackage{mathrsfs}
\usepackage[T1]{fontenc}
\usepackage{mathpazo}
\usepackage{setspace}
\usepackage{amsfonts}
\usepackage{amssymb}
\usepackage{amsmath}
\usepackage{epsfig}
\usepackage{latexsym}
\usepackage{color}
\usepackage{graphicx}
\usepackage{nicefrac}
\usepackage{pxfonts}
\usepackage[latin1]{inputenc}

\usepackage{pstricks}
\usepackage{slashed}

\date{}

\author{
  \begin{minipage}{.97\linewidth}
    \vspace{1cm}
    \begin{center}
      \begin{small}
        \textbf{P.M. Petropoulos}\footnote{marios@cpht.polytechnique.fr}, \quad
         \textbf{V. Pozzoli}\footnote{pozzoli@cpht.polytechnique.fr}\quad and \quad
  \textbf{K. Siampos\footnote{ksiampos@cpht.polytechnique.fr}}
      \end{small}
    \end{center}
    \vspace{0.5cm}
    \hspace{2cm}\begin{minipage}{.7\linewidth}
     \begin{center}
     {\it \begin{footnotesize}
     Centre de Physique Th\'eorique,
        Ecole Polytechnique, CNRS--UMR 7644\\
        91128 Palaiseau Cedex, France\\
     \end{footnotesize}}
         \end{center}
    \end{minipage}
    \vspace{0.5cm}
  \end{minipage}
}

\title{\vspace{2cm}
 \boldmath \begin{huge}
    \textbf{Self-dual gravitational instantons and geometric flows of all Bianchi types}
  \end{huge} \unboldmath
}

\newcommand{\be}{\begin{equation}}
\newcommand{\ee}{\end{equation}}

\newcommand{\bea}{\begin{eqnarray}}
\newcommand{\eea}{\end{eqnarray}}

\begin{document}

\begin{titlepage}
  \maketitle
  \thispagestyle{empty}

  \vspace{-12. cm}
  \begin{flushright}
    CPHT-RR112.1109
  \end{flushright}

  \vspace{12. cm}

  \begin{center}
    \textsc{Abstract}\\
  \end{center}

We investigate four-dimensional, self-dual gravitational instantons endowed with a product structure 
$\mathbb{R} \times\mathcal{M}_3$, where $\mathcal{M}_3$ is  {a homogeneous three-dimensional manifold} of Bianchi type. 
We analyze the general conditions under which Euclidean-time evolution in the gravitational instanton can be
 identified with a geometric flow of a metric on $\mathcal{M}_3$. This includes both unimodular and non-unimodular 
 groups, and the corresponding geometric flow is a general Ricci plus Yang--Mills flow accompanied by a diffeomorphism. 
%The identification holds for symmetric vierbeins satisfying specific commutation relations. Our findings are finally recast using the ADM decomposition and the hamiltonian formalism, which provides a natural perspective of the geometric-flow evolution underlying the instantons at hand. 

%They are  leading to a natural foliation into three-dimensional subspaces evolving in Euclidean time. For a large class of three-dimensional subspaces, the dynamics coincides with the geometric flow on the three-dimensional homogeneous slice, driven by the Ricci tensor plus an $so(3)$ gauge connection.  The metric on the three-dimensional space is related to the vielbein of the three-dimensional subspace, while the gauge field is inherited from the anti-self-dual component of the four-dimensional Levi--Civita connection.

\end{titlepage}

\onehalfspace

%\tableofcontents

%\newpage

An intriguing relationship  between three-dimensional geometric flows and 
gravitational instantons {in four dimensions has been recently investigated} in Ref. \cite{Bourliot:2009fr}. It states that self-dual solutions 
of vacuum Einstein equations  for Euclidean $\mathcal{M}_4=\mathbb{R}\times \mathcal{M}_3$ foliations, 
with homogeneous sections $\mathcal{M}_3$  of Bianchi type, are mapped onto geometric flows 
on {the three-dimensional manifold $\mathcal{M}_3$}. The geometric flows appear as a combination of Ricci and Yang--Mills flows. 

This observation, originally made in \cite{Cvetic:2001zx,Sfetsos:2006,Bakas:2006bz} for Bianchi IX, has been 
extended in  \cite{Bourliot:2009fr} for all unimodular Bianchi groups, under the assumption of diagonal metrics. 
Although for these groups any metric can eventually be taken diagonal, such a choice obscures  the reach of the 
correspondence, which might ultimately appear as a technical coincidence. Furthermore, it invalidates it for 
non-unimodular groups, where diagonal metrics are not the most general. 

The aim of the present note is to show that this correspondence holds generally, without any assumption on the 
metric and for all Bianchi classes, including the non-unimodular groups. The latter case requires the addition of
 an extra term to the Ricci part of the flow, proportional to the metric and available only in the non-unimodular 
class, as well as a prescribed diffeomorphism. Our understanding of the phenomenon at hand is now complete and 
gives confidence that it might hold similarly for Einstein gravity in higher-dimensional set-ups admitting
 self-dual solutions -- much like it does for non-relativistic gravity under the detailed-balance condition 
\cite{Horava:2008ih, Bakas:2010fm}.

{A} metric on $\mathcal{M}_4$ is generally of the following type:
\begin{equation}
\label{offdiagmet}
%\mathrm{d}s^2
g=  \mathrm{d}t^2 + g_{ij} \sigma^i  \sigma^j.
 \end{equation}
We implicitly chose a gauge with trivial shift and lapse functions. The prescribed isometry requires $ g_{ij} $ be 
a function of $t$ only, while  $\{\sigma^i, i=1,2,3\}$ are the 
left-invariant Maurer--Cartan forms of the Bianchi group. They obey
\begin{equation}
\label{conG}
\mathrm{d}\sigma^i = \frac{1}{2} c^i_{\hphantom{i}jk}\sigma^j \wedge \sigma^k.
\end{equation}
The structure constants can be put in the form (see e.g.  \cite{Ryan:1975jw})
\begin{equation}
\label{str}
c^k_{\hphantom{k}ij} = - \epsilon_{ij\ell}n^{\ell k}+\delta^k_ja_i-\delta^k_ia_j,
\end{equation}
{where $n^{\ell k}$ are the elements of a symmetric matrix $n$, and $a_i$ the components of a covector $a$. We also define the antisymmetric matrix $m$ with entries
\begin{equation}
\label{mij}
m^{ij} \equiv\epsilon^{ijk}a_k. 
\end{equation}
With these definitions, the Jacobi identity of the above algebra reads: 
\begin{equation}
\label{jacobi}
\epsilon_{ijk}m^{ij}\left(n^{k\ell}-m^{k\ell}\right)=0\Leftrightarrow
%a_k\left(n^{k\ell}-m^{k\ell}\right)=0 \Leftrightarrow 
a_k n^{k\ell} =0,
\end{equation}
whereas the trace of the structure constants is $c^j_{\hphantom{j}ij}=2a_i$.}
Unimodular groups have zero trace and
 are referred to as Bianchi A; Bianchi B are the others. Our choice for the structure constants
is presented in Tables \ref{table1} and \ref{table2}.

Self-duality conditions are naturally implemented in an orthonormal frame, where 
\begin{equation}
\label{orthoframe}
g_{ij} \sigma^i  \sigma^j =   \eta_{ij} \theta^i \theta^j,
 \end{equation}
and
\begin{equation}
\label{orthoframesym}
\eta_{ij} \theta^j=\gamma_{ij}  \sigma^j.
 \end{equation}
 Two remarks are in order here, which are at the heart of the advertised correspondence. 
Firstly, for real self-duality, only $\eta=\mathbb{1}$ or $ \mathrm{diag}(1,-1,-1)$, up to a permutation,
are allowed. This latter situation is more natural than the Euclidean one for Bianchi VIII ($SL(2,\mathbb{R})$ group) 
and for all other Bianchi groups obtained by or related to contractions of the latter (III and VI).  With this choice,
 the geometric-flow correspondence can be achieved without any complexification.  Secondly, we will not consider the most general vielbeins, but only those for which  $\gamma_{ij}$ is \emph{symmetric}.
{Although it might be restrictive}\footnote{Symmetric vielbeins are exhaustive \emph{only for} unimodular algebras. A detailed account for this issue is available in \cite{Bourliot:2009ad}, where a comprehensive and systematic analysis of all self-dual gravitational instantons of 
Bianchi type is presented in a purely Euclidean framework, without assuming symmetric vielbeins. Particular attention is drawn  to the use of symmetries 
for reducing the redundant components of the vielbein. Our perspective is here different, since we want to interpret
$\gamma_{ij}$ as a metric in order to translate its dynamics into a geometric flow but, within this assumption, we want to keep it as general as possible.}, this choice is unavoidable because $\gamma_{ij}$ appears ultimately as the metric on $ \mathcal{M}_3$,
\begin{equation}
\label{offdiagmet-flow}
 \mathrm{d}s^2=  \gamma_{ij} \sigma^i  \sigma^j,
 \end{equation}
whose appropriate flow coincides with the dynamics of the gravitational instanton. It should be stressed here, and kept in mind as an important -- and still puzzling -- feature of our analysis, that the flowing metric on  $\mathcal{M}_3$ \emph{is not} the metric on the spatial section of the corresponding gravitational instanton induced by \eqref{offdiagmet}, but rather its ``square root'' \footnote{We thank A. Petkou for drawing to our attention similar properties of scalar fields in some holographic set-ups (unpublished work).}.

In four dimensions, spin connection and curvature forms belong to the antisymmetric $\mathbf{6}$ representation of the group of local rotations $SO(4)$ (or $SO(2,2)$, when $\eta\neq\mathbb{1}$). This group factorizes as 
 $SO(3)\otimes SO(3)$ (or $SO(2,1)\otimes SO(2,1)$) and both the connection 
 %$\omega_{ab}$ 
 and the curvature 
 %$\mathcal{R}_{ab}$ $SO(4)$-valued 
 forms can be reduced 
 %with respect to the  $SO(3)_{\mathrm{(a)sd}}$ 
 as $\mathbf{6}= \mathbf{3}\otimes \mathbf{1}\oplus\mathbf{1}\otimes\mathbf{3}$,  {referred to as self-dual and anti-self-dual components. The spin-connection one-form is defined by the torsionless and metric-compatibility equations:
 \begin{equation}
\label{cartan}
\mathrm{d} \theta^a + \omega^{a\hphantom{b}}_b \wedge \theta ^b = 0,\quad 
\omega_{ab} = - \omega_{ba}. 
\end{equation}
Its decomposition in self-dual and anti-self-dual parts is} 
\begin{eqnarray}
\varsigma_i&=&\frac{1}{2}\left(\omega_{0i} + \frac{1}{2} \epsilon_{ijk}\omega^{jk}\right),\\
\alpha_i&=&
\frac{1}{2}\left(\omega_{0i} - \frac{1}{2} \epsilon_{ijk}\omega^{jk}\right).\label{redconn}
\end{eqnarray}
A similar decomposition holds for the curvature two-form.

Requiring self-duality of the curvature (see \cite{Bourliot:2009fr, Bourliot:2009ad} for details) states that the anti-self-dual Levi--Civita  $SO(3)$ (or $SO(2,1)$) connection triplet $\{\alpha_i, i=1,2,3\}$ must be a pure gauge field: $\mathrm{d} \alpha_i +\epsilon_{ijk} \alpha^j \wedge \alpha^k=0$. This is 
  %generally 
achieved with 
 \begin{equation}
\label{firoreom}
 \alpha_i
 %\frac{1}{2}\omega_{0i} - \frac{1}{2} \epsilon_{ijk}\omega^{jk} 
 =\frac{1}{2} I_ {ij}\sigma^j,
\end{equation}
where the first integral $I = \{I_{ij}\}$ %needs not be symmetric, and 
satisfies
%$ \mathrm{d} \alpha_i +\epsilon_{ijk} \alpha^j \wedge \alpha^k=0$ 
(the prime stands for $\nicefrac{\mathrm{d}}{\mathrm{d}t}$)
\begin{equation}
\label{flat-constraints}
I^{\prime}_{ij}=0,\quad
I_{i \ell}c^\ell_{\hphantom{\ell}jk}
+ \epsilon_{imn}\eta^{mp}\eta^{nq} I_{pj}
I_{qk}=0.
\end{equation}
{We can compute the spin connection using Eqs. \eqref{cartan} and our metric ansatz \eqref{offdiagmet}, \eqref{orthoframe}. Inserting its expression into \eqref{redconn} and \eqref{firoreom} we find the first-order dynamics of the vielbein components $\gamma_{ij}$ defined in \eqref{orthoframesym}. These first-order equations can be put in a compact form, useful for the subsequent developments,  by introducing a three-dimensional matrix}
\begin{equation}
\label{Om}
\Omega=\det \Gamma\left(\gamma n\gamma -\gamma m\gamma -\frac{\eta}{2}\mathrm{tr}(\gamma n \gamma \eta)\right).
\end{equation}
The matrix notation is self-explanatory: $\gamma$ and $\Gamma$ are inverse of each-other with entries $\gamma_{ij}$ and $\Gamma^{ij}$, $\eta$ stands for both $\eta_{ij}$ and $\eta^{ij}$, which are equal, $\gamma\eta=\gamma_{ij}\eta^{jk}$, \dots 
One obtains {thus} two equations: a first-order evolution equation
\begin{equation}
\label{evol}
\gamma'=-\Omega\eta\gamma - I
\end{equation}
and a constraint
\begin{equation}
\label{cons3bare}
\left[\gamma'\eta,\eta\Gamma\right]=0,
\end{equation}
equivalent, using (\ref{evol}),
to
\begin{equation}
\label{cons3}
\left[\Omega\eta+I\Gamma,\gamma\eta\right]=0.
\end{equation}
Since $\gamma$ is required to be symmetric, $\gamma'$ must also be, which imposes {through \eqref{evol}} the constraint 
\begin{equation}
\label{cons2}
\mathrm{a}(\Omega\eta\gamma+I)=0,
\end{equation}
with the notation $\mathrm{a}(P)=\nicefrac{1}{2}\left(P-P^\mathrm{T}\right)$.
Constraint \eqref{cons2} restricts the form of $\gamma$: not all symmetric vielbeins are eligible for satisfying the self-duality conditions. The precise form of $\gamma$ depends on the Bianchi class -- as well as on the basis chosen for the invariant forms (\ref{conG}), captured by the set of data $\{n,a\}$ entering (\ref{str}) (ours are displayed in Tables \ref{table1} and \ref{table2}). Notice finally that under \eqref{cons2},  constraint  \eqref{cons3} is equivalent to
\begin{equation}
\label{cons1}
\mathrm{a}(\Omega+I\Gamma\eta)=0 \Leftrightarrow \gamma m \gamma = \mathrm{a}(IM\eta),
\end{equation}
%with $\mathrm{a}(\Omega)=-\det \Gamma \gamma m \gamma$.
where $M=\mathrm{Adj}(\gamma)$ is the adjoint matrix of $\gamma$, i.e. the matrix of
the $2\times 2$ subdeterminants of~$\gamma$.

{For the reader's convenience, it is useful at this stage to outline our strategy for the next steps. The dynamics of self-dual gravitational instantons with symmetric vielbeins $\gamma_{ij}$ is governed by the evolution equation \eqref{evol}, under any two independent constraints among \eqref{cons3}, \eqref{cons2} and \eqref{cons1}. All these equations depend explicitly on the first integral $I$, which solves \eqref{flat-constraints}. We will therefore (\romannumeral1) classify the possible solutions $I$, (\romannumeral2) re-express accordingly the constraint (\ref{cons2}) as well as the evolution equation \eqref{evol}, and (\romannumeral3) interpret the resulting evolution equation as a specific geometric flow for $\gamma$ viewed as a metric on $\mathcal{M}_3$. Constraint (\ref{cons3}) (or equivalently  (\ref{cons1})) will be finally used to further  restrict the allowed form of $\gamma$.}

{Any solution of equation (\ref{flat-constraints}))} corresponds to an algebra homomorphism 
$G_3\to SO(3)$, $G_3$ being a Bianchi algebra; it can be of rank 0, 1 or 3. The general solutions are as follows:
\begin{itemize}
\item For Bianchi VIII and IX ($n=\eta, a=0$):
\begin{itemize}
\item rank 0: $I=0$,
\item rank 3: $I=\eta$ (Cayley--Hamilton theorem on $I\eta$)
\end{itemize}
In these cases, constraint (\ref{cons2}) is satisfied without restricting the form of $\gamma$.
\item In all other Bianchi classes $\det n=0$ and besides rank-0 solutions $I=0$, only rank-1 solutions exist, {for which \eqref{flat-constraints} becomes} \footnote{Equation (\ref{flat-constraints}) is strictly equivalent to $I(n+m)=\eta \mathrm{Adj}(I)^{\mathrm{T}}$  -- vanishing if $I$ is rank-1.}
\begin{equation}
\label{Ieq}
I(n+m)=0 \Leftrightarrow (n-m)I^{\mathrm{T}}=0,
\end{equation}
and they are necessarily of the form $I_{ij} = \rho_i \tau_j$. Even though  $I_{ij} $ needs not a priori be symmetric, 
the combination of constraints \eqref{cons2} and  \eqref{cons3} as well as the integrability requirement of the latter 
(validity at any time i.e. compatibility with the evolution equation \eqref{evol}) implies, after a tedious computation, 
that 
% which needs not a priori be symmetric (its antisymmetric part enters the constraint \eqref{cons2})\footnote{Explicit solutions for Bianchi III given in \cite{Bourliot:2009ad} confirm indeed that fact.}. Here, however, in order to make contact with geometric flows, we must further restrict the class of gravitational instantons and only consider those based on symmetric solutions of Eq. \eqref{flat-constraints}:
\begin{equation}
\label{conI}
I=I^{\mathrm{T}}.
\end{equation}
Under (\ref{Ieq}) and (\ref{conI}), constraint (\ref{cons2}) is equivalent to
\begin{equation}
\label{congam}
\eta \gamma(n+m)=
(n-m)\gamma\eta\Leftrightarrow [n\eta,\eta\gamma]=\{m\eta,\eta\gamma\}.
\end{equation}
%\begin{eqnarray}
%\label{congam}
%\eta \gamma(n+m)=
%(n-m)\gamma\eta&\Leftrightarrow& [n\eta,\eta\gamma]=\{m\eta,\eta\gamma\}
%\\
%&I=I^{\mathrm{T}},&
%\label{conI}
%\end{eqnarray}
This condition (also satisfied for Bianchi VIII and IX with generic $\gamma$)
% is sufficient but not necessary for (\ref{cons2}) to hold. It becomes necessary once the self-duality constraint (\ref{cons3}) is taken into account. 
 restricts the form a symmetric vielbein  $\gamma$ must have in order to be consistent with the self-duality
 evolution equation \eqref{evol}.
\end{itemize}

In all Bianchi classes, the matrix $n$ is idempotent: 
\begin{equation}
\label{idem}
n\eta n=n. 
\end{equation}
Furthermore, as a consequence of the identities relating the structure constants, one obtains
\begin{equation}
\label{netam}
n\eta m+m\eta n=m\,  \mathrm{tr}\,  n\eta. 
\end{equation}
According to this formula, the non-unimodular Bianchi algebras fall into three classes:
\begin{itemize}
\item III, VI$_{h>-1}$, VII$_{h>0}$: $n\eta m=m\eta n=m$;
\item IV: $n\eta m+m\eta n=m$;
\item V: $n\eta m=m\eta n=0$.
\end{itemize}
Hence {\eqref{idem} and  condition (\ref{congam}) imply}
\begin{equation}
\label{congamn}
n \gamma(n+m)=
(n-n\eta m)\gamma\eta\Leftrightarrow \eta \gamma(n+m \eta n)=
(n-m)\gamma n,
\end{equation}
with, as corollary, 
\begin{equation}
\label{congamncor}
n\gamma m+
m\gamma n=(m-n\eta m)\gamma\eta+ \eta\gamma(m-m\eta n).
\end{equation}

{So far we have analyzed the solutions $I$ of Eq. \eqref{flat-constraints} and processed the constraint \eqref{cons2}, with the help of various algebraic properties of the structure constants. The resulting key equations are \eqref{conI} and \eqref{congamn}, which allow for expressing the evolution equation \eqref{evol} as:}
\begin{eqnarray}
\gamma'&=&-\det \Gamma\left(\gamma n \gamma n\gamma -\frac{1}{2}\left(\gamma m \gamma n \gamma-\gamma n \gamma m \gamma\right)-\frac{\gamma}{2}\mathrm{tr}(\gamma n)^2
\right) - I\nonumber\\
\label{evolsymext}
&&-\frac{1}{2}\det \Gamma\left(
\gamma \eta\gamma(m-m\eta n)\gamma-
\gamma(m-n\eta m)\gamma \eta\gamma-
\frac{\gamma}{2}\mathrm{tr}(\gamma \eta\gamma(n\eta m-m\eta n))
\right),
\end{eqnarray}
where $I$ is {now} symmetric. This self-duality evolution equation is valid for all symmetric metrics $\gamma$ satisfying (\ref{congam}) and with $I$ being a symmetric  solution of (\ref{Ieq}),  or $I=\eta$ for rank-3 Bianchi VIII and IX. Thanks to the identity \eqref{netam},  we observe that the second line of Eq. \eqref{evolsymext} vanishes identically for all Bianchi classes but IV and V. It turns out that it also 
vanishes for those classes, {as a consequence of condition \eqref{cons3}, which has
not yet been
 taken into consideration} (we will elaborate on this in the appendix). In fact, self-duality constraint (\ref{cons3}) must be fulfilled for a solution of (\ref{evolsymext}) under (\ref{congam}) to provide a self-dual gravitational instanton.

Before analyzing the actual restrictions (\ref{cons3}) sets on the form of $\gamma$, 
% given a symmetric solution $I$ of (\ref{Ieq}), 
we would like to pause and interpret the non-vanishing part of Eq. (\ref{evolsymext}),
\begin{equation}
\label{evolsym}
\gamma'=-\det \Gamma\left(\gamma n \gamma n\gamma  -\frac{1}{2}\left(\gamma m \gamma n \gamma-\gamma n \gamma m \gamma\right)-\frac{\gamma}{2}\mathrm{tr}(\gamma n)^2
\right) - I,
\end{equation}
as a geometric flow, which is the main purpose of the present note. This was achieved in \cite{Bourliot:2009fr} for Bianchi A (non-unimodular) under the assumption of diagonal $\gamma$. The geometric flow was shown to be a Ricci flow combined with a Yang--Mills flow produced by a flat, non-flowing and diagonal $SO(3)$ Yang--Mills connection on $\mathcal{M}_3$. The origin of the {Yang--Mills connection}  was the flat anti-self-dual part of the Levi--Civita connection on $\mathcal{M}_4$, appearing as the first integral (\ref{firoreom}). 

For general metrics $\gamma$ (see (\ref{offdiagmet-flow})) on $G_3$-left-invariant $\mathcal{M}_3 $, the Ricci tensor reads \footnote{The scalar curvature is 
$
S[\gamma]=\mathrm{tr}(\Gamma R[\gamma])=\mathrm{tr}(\Gamma N)-\frac{\det \Gamma}{2} \mathrm{tr}(\gamma n)^2 -5 a\Gamma a
$.}:
\begin{equation}
\label{Ricci}
R[\gamma]=N+\det \Gamma\left(\gamma n \gamma n\gamma -\gamma m \gamma n \gamma+\gamma n \gamma m \gamma-\frac{\gamma}{2}\mathrm{tr}(\gamma n)^2
\right) +a\otimes a -2\gamma\,  a\Gamma a.
\end{equation}
In the last term, $a\Gamma a$ stands for $a_i\Gamma^{ij}a_j$, while $N$ is the Cartan--Killing metric of the $G_3$ algebra: 
\begin{equation}
\label{CK}{
N_{ij}= -\frac{1}{2}  \epsilon_{\ell i m}
\epsilon_{kjn}n^{mk}n^{n \ell} -a_ia_j.}
\end{equation}
The appropriate Yang--Mills connection to consider here is $SO(2,1)$ for Bianchi VI or VIII,  and $SO(3)$ otherwise, since it reflects, in each case, the four-dimensional anti-self-dual Levi--Civita connection:
 $A\equiv A_i\sigma^i=-\lambda_{ij}T^j \sigma^i$ with $\left[T_i,T_i\right]=-\epsilon_{ijk}T^k$. As usual $T^i=\eta^{ij}T_j$ with $\eta=\mathbb{1}$ for $SO(3)$  and $\mathrm{diag}(1,-1,-1)$ for  $SO(2,1)$.  The absence of  flow for the Yang--Mills connection states that $A^{\prime}=0$, while flatness requires 
\begin{equation}
\label{flat-constraints-YM}
F = \mathrm{d}A +[ A, A]\equiv 0
 \Leftrightarrow    \lambda_{i \ell}c^\ell_{\hphantom{\ell}jk}
+ \epsilon_{ijk} \eta^{jm} \eta^{kn}\lambda_{m j}
\lambda_{n k}=0.
\end{equation}
It is straightforward to show\footnote{Notice also the normalisation of the generators: $\mathrm{tr}(T_iT_j)=-2\eta_{ij}$.} that $-\nicefrac{1}{2}\,\mathrm{tr}(A_iA_j)=\eta^{k\ell}\lambda_{ki}\lambda_{\ell j}$. Combining the latter with \eqref{Ricci}, we conclude that  Eq. (\ref{evolsym}) is recast as
\footnote{Here $\mathrm{s}(\nabla a)$ stands for the symmetric part of $\nabla a$: $\mathrm{s}(\nabla a)=\frac{\det \Gamma}{2}\left(\gamma n \gamma m \gamma-\gamma m \gamma n \gamma\right)+a\otimes a -\gamma\,  a\Gamma a $.}
\begin{equation}
\label{Ric}
\frac{\mathrm{d}\gamma}{\mathrm{d}t}=-R[\gamma] +\mathrm{s}(\nabla a)  - \gamma \, a\Gamma a-\frac{1}{2}\mathrm{tr} \left( A\otimes  A\right).
\end{equation}
 This describes a geometric flow driven by the Ricci tensor, combined with Yang--Mills as well as a diffeomorphism generated by $a$ and an invariant component of the scalar curvature. The matching requires the following  relationship to hold between the flat anti-self-dual Levi--Civita connection $I_{ij}$ and the flat Yang--Mills connection $\lambda_{ij}$:
\begin{equation}
\label{corr}
N-I
=
\lambda^{\mathrm{T}}\eta\lambda, 
\end{equation}
where $I$ (symmetric) and $\lambda$ satisfy \eqref{flat-constraints} and \eqref{flat-constraints-YM} respectively. This equation  is indeed consistent:
\begin{itemize}
\item For Bianchi VIII and IX, $N=\eta$, $I=0$ or $\eta$ and $\lambda=\eta$ or $0$. Thus \eqref{corr} translates onto $I+\lambda=\eta$.
\item For all other types, Eqs. \eqref{flat-constraints} and  \eqref{flat-constraints-YM} are equivalent to $I(n+m)=(n-m)I=0$ and $\lambda(n+m)=(n-m)\lambda^{\mathrm{T}}=0$ respectively. Since
$N(n+m)=(n-m)N=0$ (as a consequence of Jacobi identity {\eqref{jacobi}}),  any rank-0 or rank-1 solution $I$ provides, through \eqref{corr}, a solution for $\lambda$ and vice-versa.
\end{itemize}
It should finally be stressed that the consistency condition \eqref{conI} was instrumental in reaching 
 \eqref{congamn} from  \eqref{cons2}, and therefore in rewriting \eqref{evol} as \eqref{evolsym}, and further as \eqref{Ric} (for Bianchi IV and V, constraint \eqref{cons3} was also necessary to ensure the equivalence of \eqref{evolsymext} and  \eqref{evolsym}  -- see the appendix).
 
The above demonstrates the advertised general correspondence between gravitational instantons and geometric flows for metrics $\gamma$ satisfying \eqref{congam} (and \eqref{cons3} for Bianchi IV and V), and first integral $I$ satisfying \eqref{conI}. Two questions are in order at this stage. The first concerns the self-consistency of the geometric flow \eqref{Ric}. Within the framework of self-dual gravitational instantons, the metrics $\gamma$ are forced to fulfill  \eqref{congam} (and \eqref{cons3} for Bianchi IV and V).  Is the flow evolution compatible with this constraint? The answer is positive and one easily  checks that $\gamma'$ satisfies \eqref{congam} if $\gamma$ does. The second question is whether any consistent geometric flow of this type is eligible as gravitational instanton. The answer in this case in negative, because only $\gamma$s further restricted to \eqref{cons3}, constraint that has not yet been taken explicitly into account, can be promoted to four-dimensional self-dual solutions.  

\begin{table}
\caption{Basis of invariant forms and restrictions on $\gamma$ -- unimodular groups}
\label{table1}
\begin{tabular}[c]{|c|c|c|c|}
\hline
Type &  $n,\eta,a=0$ & Restrictions from \eqref{cons2} & Restrictions from \eqref{cons3} \\
\hline
\hline
I  & $n=0\quad \eta =\mathbb 1$& none  & none rank-1  \\
\hline
II  &  $n = \mathrm{diag} (0,1,0)$ & $\gamma_{12}=0=\gamma_{23}$ & $\gamma_{12} =0= \gamma_{23}$:\ rank-0  \\
&$\eta =\mathbb{1}$& &$\gamma_{13}=0$ and/or $\gamma_{11}=\gamma_{33}$:\ rank-1   \\\hline 
VIII &  $n =\eta = \mathrm{diag} (1,-1,-1) $ & none & none rank-0 \\
&  & &none rank-3: $I=\eta$  \\
\hline
IX &  $n=\eta =\mathbb 1$& none& none rank-0  \\
&    & &none rank-3: $I=\eta$  \\
\hline
VI$_{-1}$ &  $n=\mathrm{diag} (0,1,-1)$ &$\gamma_{12}=0=\gamma_{13}$ & $\gamma_{12} =0= \gamma_{13}$  \\
&  $\eta = \mathrm{diag} (-1,1,-1) $ & &rank-0,1  \\
\hline
VII$_0$ & $n=\mathrm{diag}(1,1,0)$ & $\gamma_{12}=0=\gamma_{23}$ & $\gamma_{13} =0 =\gamma_{23}$  \\
&   $\eta = \mathrm{diag} (1,1,1) $ & &rank-0,1  \\
\hline
\end{tabular}
\end{table}

The self-duality constraint  \eqref{cons3} does not affect Bianchi VIII or IX, for which $\gamma$ remains generic. It does however further restrict $\gamma$ in the other Bianchi groups, without altering the consistency of the flow equation  \eqref{evolsym}  (this follows immediately from the original formulation of the self-duality  constraint \eqref{cons3bare}).
For the non-unimodular Bianchi class, \eqref{cons3} or equivalently \eqref{cons1}, imposes that all rank-0 ($I=0$) metrics must have $\det \gamma=0$. This corresponds to singular gravitational instantons or degenerate geometric flows. The same holds for rank-1 ($I\neq 0$) except for  Bianchi III, which admits non-singular self-dual gravitational instantons corresponding to regular geometric flows. These properties are collected in the appendix; more details on the analysis of \eqref{cons3} can be found in \cite{Bourliot:2009ad} for general vielbeins (as opposed to the symmetric ones used here)\footnote{In comparing the the present developments with those of \cite{Bourliot:2009ad}, the reader should bear in mind that, besides the absence of symmetry restriction for the vielbein, the signature used in that reference was Euclidean,  whereas here it is Euclidean for Bianchi I, II, IV, V, VII, IX and ultra-hyperbolic for III, VI and VIII.}. For the reader's convenience, the case of Bianchi III will be presented in the appendix, whereas we summarize the results for the other classes
%\footnote{This is valid within our choice of basis for the left-invariant forms.}  
in Tables \ref{table1} and \ref{table2}. There, the forms of symmetric $\gamma$s complying with 
\eqref{congam}  are displayed, together with their restriction following the self-duality constraint \eqref{cons3}. 
These expressions depend on the basis of invariant forms, which are also specified.

The above developments conclude on the advertised correspondence among gravitational instantons and geometric flows. 
It would be interesting to {provide a satisfactory  geometrical interpretation of the ``square root''  of $g$ as the flowing metric on $\mathcal{M}_3$,  and to}
understand how the first-order self-duality equations \eqref{evol} could emerge directly 
from the action -- as they do for $g$  in the non-relativistic set-up under detailed balance  \cite{Horava:2008ih, Bakas:2010fm} -- 
 following the split formalism of Refs. \cite{Mansi:2008br, Mansi:2008bs}. The analysis of general geometric flows of the 
type \eqref{Ric}, satisfying \eqref{congam} but not \eqref{cons3},  i.e. beyond those which are interpreted as self-dual
 gravitational instantons, is also an open and challenging problem. Interesting issues such as the existence of entropy functionals, or universality properties in the large-time behaviour deserve further investigation. 

Let us finally stress the specificity of four dimensions in respect to the relationship between geometric flows and gravitational
instantons. Even though self-duality can be imposed in other dimensions (such as 7 or 8 thanks to the $G_2$ structure and 
to the octonions \cite{Corrigan:1982th}), the holonomy group has not the factorization property of $SO(4)$ or $SO(3,1)$, 
which was instrumental here. Although not excluded, it seems more difficult to find a class of instantons that could be interpreted as geometric flows of a lower-dimensional geometry.     

\begin{table}
\caption{Basis of invariant forms and restrictions on $\gamma$ -- non-unimodular groups}
\label{table2}
\begin{tabular}[c]{|c|c|c|c|}
\hline
Type &  $n,\eta,a$ & Restrictions from \eqref{cons2} & Restrictions from \eqref{cons3} \\
\hline
\hline
III &  $n=\mathrm{diag}(0,1,-1)\quad a =(1,0,0)$ & $\gamma_{12}=\gamma_{13}$ & $\gamma_{11}, \gamma_{13}$ given in \eqref{rel}\\
&   $\eta = \mathrm{diag} (-1,1,-1)$ & $\gamma_{22}=\gamma_{33}$ & \\
\hline
IV &  $n=\mathrm{diag}(0,1,0)$ & $\gamma_{12}=0=\gamma_{13}$, & singular from \eqref{gAg} \\
&  $\eta = \mathbb 1 \quad a=(1,0,0)$ &$\gamma_{23}=\gamma_{22}+\gamma_{33}$ & $\gamma_{22} = 0 = \gamma_{33}$ \\
\hline
V & $n=0 \quad \eta = \mathbb 1$ & $\gamma_{12} =0= \gamma_{13} $ &singular from \eqref{gAg} \\
&  $\quad a=(1,0,0)$ & $\gamma_{22}=-\gamma_{33}$ & $\gamma_{22} = 0 = \gamma_{23}$ \\
\hline
VI$_{h>-1}$ & $n=\mathrm{diag}(0,1,-1)\quad a=(h+1,0,0)$ & $\gamma_{12} = 0= \gamma_{13} $  & singular from \eqref{gAg}\\
&  $\eta = \mathrm{diag}(-1,1,-1)$  & $\gamma_{22}=\gamma_{33}$ & $\gamma_{22}=\gamma_{23}$ 
\\
\hline
VII$_{h>0}$  & $n=\mathrm{diag}(1,1,0) \quad \eta= \mathbb 1$ & $\gamma_{13} = 0 = \gamma_{23}$ & singular from \eqref{gAg}\\
&  $ a =(0,0,h)$ &$\gamma_{11}=-\gamma_{22}$ & $\gamma_{11} = 0 = \gamma_{12}$ \\
\hline
\end{tabular}
\end{table}

\section*{Acknowledgements}

The authors would like to thank I. Bakas, F. Bourliot, J. Estes, N. Karaiskos, Ph. Spindel
and K. Sfetsos for stimulating discussions and acknowledge financial support by  the ERC Advanced Grant 226371, 
the IFCPAR programme 4104-2 and the ANR programme blanc NT09-573739.
V.~Pozzoli and K. Siampos were supported by the the ITN programme PITN-GA-2009-237920.
K. Siampos  acknowledges hospitality and financial support  of the Theory Unit of CERN.
P.M. Petropoulos and K. Siampos would like to thank the University of Patras for hospitality.

%\newpage

\appendix
\section*{On non-unimodular groups}\label{IVV}

For metrics $\gamma$ obeying \eqref{cons2}, the self-duality evolution equation \eqref{evol} translates into \eqref{evolsymext}. The second line of  this equation vanishes identically in all Bianchi classes for which $n\eta m = m \eta n = m$. This excludes IV and V (see \eqref{netam}). For the latter classes, it turns out that the terms at hand vanish provided $\gamma$ is subject to the self-duality constraint \eqref{cons3} or equivalently \eqref{cons1}. This statement is based on the simple fact that for all groups but Bianchi III, 
\begin{equation}
\label{gAg}
\gamma m \gamma=0,
\end{equation}
as a consequence of \eqref{cons1} in combination with \eqref{cons3}. Proving that the second line of \eqref{evolsymext} vanishes is thus a matter of simple algebra with the help of expressions \eqref{congam}--\eqref{congamncor}. 

The proof of \eqref{gAg} goes as follows. 
For all non-unimodular groups, a  first integral $I$ solving \eqref{Ieq} is either vanishing or rank-one. In the first case, \eqref{cons1} demonstrates \eqref{gAg}. In the second, the general solution for symmetric $I$ is
\begin{equation}
\label{I-III}
I_{ij}=\kappa a_i a_j,
\end{equation}
($\kappa$ is an arbitrary constant) for Bianchi IV, V, VI \& VII. The constraint \eqref{cons1}, written in Poincar\'e-dual form and  combined with \eqref{cons2} reads:
 \begin{equation}
\label{mnm}
\frac{2}{\kappa} M^{i\ell}a_\ell=n^{ij}\eta_{jk}M^{k\ell}a_\ell,
\end{equation}
where $M^{i\ell}$ are the entries of %the $2\times 2$ subdeterminant matrix 
$\mathrm{Adj}(\gamma)$. From Eq. \eqref{mnm}, we observe that $M^{i\ell}a_\ell$  are the components of an eigenvector of $n\eta $ 
 with eigenvalue $\nicefrac{2}{\kappa}$. Multiplying iteratively \eqref{mnm} by $n\eta$ from the left and using \eqref{idem}, we conclude that the eigenvalue of the eigenvector at hand is an arbitrary natural power of $\nicefrac{2}{\kappa}$. This is possible only if $M^{i\ell}a_\ell$ vanish. Since 
  \begin{equation}
\label{gag-dual}
\epsilon^{ijk} (\gamma m \gamma)_{ij} = 2M^{k\ell} a_\ell,
\end{equation}
\eqref{gAg} is proven in full generality, without reference to any particular choice of basis.  

Notice  that following \eqref{gAg}, $a_j$ is an eigenvector of $M^{ij}$ with zero eigenvalue. Therefore $\det \gamma= 0$, as already announced, for all self-dual gravitational instantons based on non-unimodular Bianchi groups, except for III. 
 
%\section{The case of Bianchi III}\label{III}

For Bianchi III the above do not  hold because \eqref{I-III} does not provide the most general symmetric solution of \eqref{Ieq}.
%\footnote{The most general solution is $I=\kappa \rho\otimes \rho$ with $\rho_1$ arbitrary and $\rho_2=\rho_3$ in our basis.}.
The generic first integral $I$ satisfying \eqref{Ieq} and \eqref{conI} is instead
\begin{equation}
\label{IIIfi}
I=\begin{pmatrix}
\mu&\chi&\chi\\
\chi&-\nu&-\nu\\
\chi&-\nu&-\nu
\end{pmatrix},\quad \chi^2+\mu\nu=0.
\end{equation}
Bianchi III is the only non-unimodular case admitting non-degenerate $\gamma$s, once all constraints \eqref{cons3} and \eqref{cons2} are taken into account. 
The consistency of symmetric $\gamma$, Eq. \eqref{congamn}, sets
\begin{equation}
\label{IIIcongamn}
\gamma_{12}=\gamma_{13},\quad \gamma_{22}=\gamma_{33},
\end{equation}
whereas using \eqref{IIIfi}, the
evolution equation \eqref{evolsym} matches the geometric-flow equation \eqref{Ric} provided $\lambda$ satisfies \eqref{flat-constraints-YM} and \eqref{corr}. The general solution of  \eqref{flat-constraints-YM} is
\begin{equation}
\label{IIIgauge}
\lambda=\begin{pmatrix}
\rho_1 \lambda&\rho_1\zeta&\rho_1\zeta\\
\rho_2 \lambda&\rho_2\zeta&\rho_2\zeta\\
\rho_3 \lambda&\rho_3\zeta&\rho_3\zeta
\end{pmatrix}.
\end{equation}
Requiring \eqref{corr} with \eqref{IIIfi} leads to the following set of equations, 
\begin{equation}
\left(\rho_1\lambda\right)^2 = 2+\mu,\quad 
\left(\rho_1^2-\rho_2^2+\rho_3^2\right)\lambda \zeta= \chi,\quad 
\left(\rho_1^2-\rho_2^2+\rho_3^2\right)\zeta^2= -\nu,
\end{equation}
which always admit a solution, either for first integral $I$ of rank-0 ($\mu=\nu=\chi=0$), or of rank-1 (either $\mu, \nu$ or $\chi$ non-zero) .

Self-duality also demands \eqref{cons3} to be satisfied. The further constraints on $\gamma$ are
\begin{equation}
\label{rel}
\gamma_{11}=\frac{\chi^2+2\nu}{2\nu^2}\left(\gamma_{23}+\gamma_{33}\right),\quad 
\gamma_{13}=-\frac{\chi}{2\nu}\left(\gamma_{23}+\gamma_{33}\right)
 \end{equation}
and the general solution of the evolution equation \eqref{evolsym} reads:
\begin{equation}
\begin{split}
\gamma_{11} (t) &= (1 - \frac{\mu}{2 } )\ t+ \gamma_{11} (0), \\
\gamma_{13} (t) &= - \frac{\chi}{2} \ t + \gamma_{13} (0),\\
\gamma_{33} (t) &= \gamma_{33}(0) \left(1+ \frac{\nu t }{\gamma_{33} (0) + \gamma_{23} (0)} \right), \\
\gamma_{23} (t) &= \gamma_{23}(0) \left( 1 + \frac{\nu t}{\gamma_{33} (0) + \gamma_{23} (0)} \right),
\end{split}
\end{equation}
where the initial conditions are related by \eqref{rel}, and $\mu,\nu,\chi$ constrained by \eqref{IIIfi}.

Self-dual Bianchi-III gravitational instantons of the above type were analyzed in \cite{Bourliot:2009ad}, for the general case of non-symmetric vielbein. Among others, they exhibit
naked singularities (points where Kretschmann's invariant becomes infinite). From the viewpoint of the geometric flow, $\gamma$ is a metric on $\mathcal{M}_3$  evolving under \eqref{Ric}. Its components are linearly expanding or shrinking, depending on the parameters $\mu$ and $\nu$, and on the initial values $\gamma_{23}(0)$ and $\gamma_{33}(0)$. In particular, the scalar curvature depends on time as
\begin{equation}
\label{IIIsccu}
S= \frac{8\nu}{\gamma_{33}(0)+ \gamma_{23}(0) +\nu t},
\end{equation}
and can describe the relaxation of a singular configuration toward flatness, or the appearance of a singularity from an ancient flat space. Any attempt to go deeper in this analysis requires a more general study of flows of the type \eqref{Ric}, which stands beyond our present motivation.


\begin{thebibliography}{99}

%\cite{Bourliot:2009fr}
\bibitem{Bourliot:2009fr}
  F.~Bourliot, J.~Estes, P.M.~Petropoulos and Ph.~Spindel,
 ``Gravitational instantons, self-duality and geometric flows,''
  Phys.\ Rev.\  {\bf D81} (2010) 104001
, \href{http://arxiv.org/abs/0906.4558} {[arxiv 09064558 [hep-th]]}.
  %%CITATION = PHRVA,D81,104001;%%
  

\bibitem{Cvetic:2001zx}
  M.~Cveti\v{c}, G.W.~Gibbons, H.~Lu and C.N.~Pope,
  ``Cohomogeneity-one manifolds of Spin(7) and G(2) holonomy,''
  Phys.\ Rev.\   {\bf D65} (2002) 106004
,\href{http://arxiv.org/abs/hep-th/0108245}{[arXiv:hep-th/0108245]}.
  %%CITATION = PHRVA,D65,106004;%%

  \bibitem{Sfetsos:2006} K. Sfetsos, unpublished work,
\href{http://www.cc.uoa.gr/~papost/SFETSOS.pdf}{http://www.cc.uoa.gr/$\sim$papost/SFETSOS.pdf}

    \bibitem{Bakas:2006bz}
  I.~Bakas, D.~Orlando and P.M.~Petropoulos,
  ``Ricci flows and expansion in axion-dilaton cosmology,''
  JHEP {\bf 0701} (2007) 040
,\href{http://arxiv.org/abs/hep-th/0610281}{[arXiv:hep-th/0610281]}. 
  %%CITATION = JHEPA,0701,040;%%

  \bibitem{Horava:2008ih}
  P.~Ho\v{r}ava,
  ``Membranes at quantum criticality,''
  JHEP {\bf 0903}, 020 (2009)
  \href{http://arxiv.org/abs/0812.4287}{[arXiv:0812.4287 [hep-th]]}.
  %%CITATION = JHEPA,0903,020;%%
  
  
  %\cite{Bakas:2010fm}
\bibitem{Bakas:2010fm}
  I.~Bakas, F.~Bourliot, D.~L\"ust and P.M.~Petropoulos,
  ``Geometric flows in Ho\v{r}ava--Lifshitz gravity,''
  JHEP {\bf 1004} (2010) 131
\href{http://arxiv.org/abs/1002.0062}{[arXiv:1002.0062 [hep-th]]}.
  %%CITATION = JHEPA,1004,131;%%
  
  \bibitem{Ryan:1975jw}
  M.P.~Ryan and L.C.~Shepley,
  ``Homogeneous relativistic cosmologies,''
%\href{http://www.slac.stanford.edu/spires/find/hep/www?irn=1442376}{SPIRES entry}
{\it  Princeton, Usa: Univ. Pr. (1975)   Princeton Series In Physics}
  
  %\cite{Bourliot:2009ad}
\bibitem{Bourliot:2009ad}
  F.~Bourliot, J.~Estes, P.M.~Petropoulos and Ph.~Spindel,
  ``$G3$-homogeneous gravitational instantons,''
  Class.\ Quant.\ Grav.\  {\bf 27} (2010) 105007
\href{http://arxiv.org/abs/0912.4848}{[arXiv:0912.4848 [hep-th]]}.
  %%CITATION = CQGRD,27,105007;%%

 \bibitem{Mansi:2008br}
  D.S.~Mansi, A.C.~Petkou and G.~Tagliabue,
  ``Gravity in the $3+1$-split formalism I: holography as an initial value problem,''
  Class.\ Quant.\ Grav.\  {\bf 26} (2009) 045008
\href{http://arxiv.org/abs/0808.1212}{[arXiv:0808.1212 [hep-th]]}.
  %%CITATION = CQGRD,26,045008;%%

  \bibitem{Mansi:2008bs}
  D.S.~Mansi, A.C.~Petkou and G.~Tagliabue,
  ``Gravity in the $3+1$-split formalism II: self-duality and the emergence of the gravitational Chern--Simons in the boundary,''
  Class.\ Quant.\ Grav.\  {\bf 26} (2009) 045009
\href{http://arxiv.org/abs/0808.1213}{[arXiv:0808.1213 [hep-th]]}.
  %%CITATION = CQGRD,26,045009;%%

%\cite{Corrigan:1982th}
\bibitem{Corrigan:1982th}
  E.~Corrigan, C.~Devchand, D.B.~Fairlie and J.~Nuyts,
  ``First-order equations for gauge fields in spaces of dimension greater than four,''
\href{http://www.sciencedirect.com/science/article/pii/0550321383902444}{Nucl.\ Phys.\  {\bf B214} (1983) 452}.
  %%CITATION = NUPHA,B214,452;%%




\end{thebibliography}
\end{document}